\documentclass[12pt]{article}
\usepackage[english]{babel}
\usepackage{amsmath}
\usepackage{amssymb}
\usepackage{bbm}
\usepackage{color}
\usepackage[titletoc,toc,title]{appendix}
\Roman{section}
\newcommand{\eq}[1]{\begin{equation}#1\end{equation}}
\newcommand{\naw}[1]{\left(#1\right)}
\newcommand{\ket}[1]{\left|#1\right>}
\newcommand{\bra}[1]{\left<#1\right|}
\newcommand{\av}[1]{\left<#1\right>}

\newcommand{\modu}[1]{\left|#1\right|}

\begin{document}

\begin{center}
\textsc{\Large{Note on maximally entangled Eisert-Lewenstein-Wilkens quantum games}}
\newline

\large{Katarzyna Bolonek-Laso\'n}\footnote{kbolonek@uni.lodz.pl}\\ 
\emph{\normalsize{Faculty of Economics and Sociology, Department of Statistical Methods, \\University of Lodz,
41/43 Rewolucji 1905 St., 90-214 Lodz,  Poland.}}\\
\large{Piotr Kosi\'nski}\footnote{pkosinsk@uni.lodz.pl}\\
\emph{\normalsize{Department of Computer Science, Faculty of Physics and Applied Informatics, University of Lodz, 149/153 Pomorska St., 90-236 Lodz, Poland.}}
\end{center}

\begin{abstract}
Maximally entangled Eisert-Lewenstein-Wilkens games are analyzed. For a general class of gates defined in the previous papers of the first author the general conditions are derived which allow to determine the form of gate  leading to maximally entangled games. The construction becomes particularly simple provided one does distinguish between games differing by relabelling of strategies. Some examples are presented.
\end{abstract}

\section{Introduction}
The seminal papers of Eisert, Lewenstein and Wilkens (ELW) \cite{EisertWL}, \cite{EisertW} opened new field of intensive research called the theory of quantum games \cite{Meyer}$\div$\cite{Nawaz6}. They proposed a method of constructing a quantum counterpart of a given non-cooperative classical game.

Eisert, Lewenstein and Wilkens pointed out that there is an intimate connection between the theory of quantum games and quantum communication. They speculated also that games of survival are being played on molecular level ruled by the laws of quantum mechanics.

Their approach constitutes one of the paradigms of the theory of quantum games. This is because it provides simple yet subtle scheme which allows to study the influence of non-classical correlations which break Bell-like inequalities on the properties (in particular the efficiency in the presence of restricted resources) of classical games. In fact, the ELW game may be viewed as a strightforward generalization of classical symmetric noncooperative game where the players take advantage from the fact that quantum probability distributions violate, in general, such inequalities.

The original proposal concerned the quantization of symmetric 2-players  game with two strategies at each player's disposal. It can be generalized to the case of arbitrary number $N$ of admissible strategies. For general $N$ the structure of the game becomes much richer. The key role in construction of quantum game is played by the gate  which allows quantum correlations to influence the outcome of the game. In the original ELW proposal the gate  depends on one arbitrary parameter. For $N$-strategies games one can construct the gate depending on $\binom{N}{2}$ parameters \cite{Bolonek3}$\div$\cite{Bolonek5}.

The strength of quantum correlations is measured by the quantum entanglement. It is not surprising that the so called maximally entangled games play a distinguished role. They do not admit nontrivial pure Nash equilibria \cite{BenjaminHay}, \cite{Landsburg1}, \cite{Bolonek} and exhibit additional structures \cite{Landsburg1}, \cite{Bolonek}. They are particularly worth of being studied in more detail.

Whether the game is maximally entangled or not depends on the choice of the gate. In the present paper we derive the general conditions on the parameters entering the gates introduced in Refs. \cite{Bolonek3}$\div$\cite{Bolonek5} for the game to be maximally entangled. They appear to be quite straightforward and manageable. What is also important they are applicable to the family of gates  which seem to exhaust all interesting cases and which provide a natural generalization of the gate introduced by Eisert et al.

The paper is organized as follows. In sec. II we remind the definition of the general ELW game. Then, in Sec. III we derive the conditions which must be imposed on gate to yield maximally entangled game. Sec. IV is devoted to some examples. In Sec. V we show that the relevant gate operators can be easily written out provided we distinguish between the games differing only by relabelling of Alice and Bob strategies. Fianally, Sec. VI contains some concluding remarks. Brief information on Cartan subalgebras is presented in Appendix.      

\section{Quantum ELW game}
Let us describe the general setting for the quantum ELW game. The starting point is a classical two-players symmetric game. Each player has $N$ strategies at his/her disposal (the original Eisert et al. proposal corresponds to $N=2$). The classical game is completely defined by $N\times N$ payoff matrices $P^{A,B}$ with matrix elements $P^{A,B}_{\sigma,\sigma'}$, $\sigma,\sigma'=1,...,N$; $P^A_{\sigma,\sigma'}$ ($P^B_{\sigma,\sigma'}$) denotes Alice (Bob) payoff in the case Alice and Bob choose the strategies $\sigma$ and $\sigma'$, respectively.

In order to construct a quantum counterpart of the game we ascribe to any player a $N$-dimensional Hilbert space $\mathcal{H}$ spanned by the vectors
\begin{equation}
\ket{1}=\left( \begin{array}{c}
1\\
0\\
\vdots\\
0
\end{array}\right),\quad\cdots,\quad \ket{N}=\left( \begin{array}{c}
0\\
0\\
\vdots\\
1
\end{array}\right).\label{ab} 
\end{equation}
The Hilbert space of the game is $\mathcal{H\otimes\mathcal{H}}$. We start with the vector $\ket{1}\otimes\ket{1}$. The key element of the definition of quantum game is the choice of an unitary operator (the gate) $J$ which introduces quantum entanglement. The initial state of the game is defined as
\begin{equation}
\ket{\Psi_i}=J\naw{\ket{1}\otimes\ket{1}}.\label{a}
\end{equation}
Before proceeding further let us comment briefly on the choice of initial state. One of the main points in construction of ELW game is that any entanglement is introduced and controlled by the gate $J$. Therefore, we start with an unentangled vector (i.e. with Schmidt number 1) which can be written as tensor product of Alice and Bob states; so, in general,
\eq{\ket{\Psi_i}=J\naw{\ket{\Psi}\otimes\ket{\Psi}}\label{da1},}
$\ket{\Psi}\in \mathcal{H}$ being an arbitrary normalizable state. Now, there exists an unitary operator $U$ such that 
\eq{\ket{\Psi}=U\ket{1}.\label{da2}}
Note that the phase of $\ket{\Psi}$ is irrelevant (cf. eq. (\ref{da}) below) so it can be adjusted such that $U\in SU\naw{N}$. Moreover, $U$ is not unique; in fact, it can be multiplied from the right by any element of $S\naw{U(1)\times U(N-1)}$. By virtue of eqs. (\ref{da1}) and (\ref{da2}) $\ket{\Psi}$ may be always replaced by $\ket{1}$ provided simultaneously the replacements $J\rightarrow\naw{U\otimes U }J\naw{U^+\otimes U^+}$, $U_A\rightarrow UU_AU^+$, $U_B\rightarrow UU_BU^+$. So $J$ changes by a local unitary transformation and the strategies are relabelled; we conclude that one can take $\ket{\Psi}=\ket{1}$ without loosing generality.

Once the initial state is defined Alice and Bob perform their moves represented by unitary matrices $U_{A,B}\in SU(N)$. Then the final measurement is made which yields the final state
\begin{equation}
\ket{\Psi_f}=J^+\naw{U_A\otimes U_B}J\naw{\ket{1}\otimes\ket{1}}.
\end{equation}
This allows us to compute the players expected payoffs
\begin{equation}
\$^{A,B}=\sum_{\sigma,\sigma'=1}^{N}P^{A,B}_{\sigma,\sigma'}\modu{\av{\sigma,\sigma'|\Psi_f}}^2\label{da}
\end{equation}
where $\ket{\sigma,\sigma'}\equiv\ket{\sigma}\otimes\ket{\sigma'}$.

As it was mentioned above it is the form of the gate $J$ which determines the properties of the game. If $J$ is trivial (i.e. a product of two local unitary matrices) the game reduces to the classical one. Any pure quantum strategy corresponds to some, in general mixed, classical one (actually, there is some overcounting, i.e. a number of pure quantum strategies correspond to the same classical strategy). The situation changes if $J$ ceases to be the product of local unitary operators. Then the initial state of the game becomes entangled. The degree of entanglement plays a crucial role in the study of game properties. Roughly speaking the more entangled the state is the more the outcome probabilities differ from the values allowed by inequalities of Bell type resulting in properties not shared by the classical game one starts with. In view of this it is natural to ask about the maximally entangled case. As we have already mentioned in the Introduction maximally entangled games are distinguished by their properties. One of us has shown \cite{Bolonek5} that the key element here is that the stability subgroup of initial state, i.e. the subgroup of the group $SU(N)\times SU(N)$ of all strategies which consists of elements which leave invariant the initial state is isomorphic to the diagonal subgroup of $SU(N)\times SU(N)$. This has, for example, deep impact on the structure of Nash equilibria. It is known \cite{Glicksberg}, \cite{Lee} that the validity of Nash theorem extends to the quantum domain; as in the classical games the Nash equilibria correspond, in general, to mixed strategies. However, for maximally entangled case one finds an additional property: even if the classical payoff matrix admits pure Nash equilibria, they cease to exist in the quantized version (except some trivial cases). Moreover, the study of mixed strategies corresponding to Nash equilibria simplifies considerably. This is due to the fact that, again as a consequence of the form of stability subgroup, the final outcome depends essentially only on the product of the matrices representing players moves \cite{Bolonek5}. This property has been successfully used in the case $N=2$ to provide the classification of mixed Nash equilibria \cite{Landsburg1}, \cite{Landsburg2}.

In order to construct the gate operator $J$ we assume that the quantum game is still symetric and all classical pure strategies are contained in the set of pure quantum ones. It has been shown in Refs. \cite{Bolonek3}, \cite{Bolonek5} that one can construct a multiparameter family of gates $J$ obeying this condition. To this end we define first the unitary matrix $V$ by \cite{Bolonek3}
\begin{equation}
V_{\alpha,\beta}=\frac{1}{\sqrt{N}}\overline{\varepsilon}^{\naw{\alpha-1}\naw{\beta-1}},\qquad \alpha,\beta=1, ...,N
\end{equation}
where $\varepsilon=\exp\naw{\frac{2i\pi}{N}}$ is the first primitive root from unity. Let $\Lambda_i$, $i=1,...,N-1$, be any basis in the Cartan subalgebra of $SU(N)$ ( consisting of diagonal traceless hermitean matrices). Define
\begin{equation}
\widetilde{J}=\exp\naw{i\sum_{k=1}^{N-1}\lambda_k\naw{\Lambda_k\otimes\Lambda_k}+\frac{i}{2}\sum_{k\neq l=1}^{N-1}\mu_{kl}\naw{\Lambda_k\otimes\Lambda_l+\Lambda_l\otimes\Lambda_k}}\label{aa}
\end{equation}
with $\lambda_k$, $\mu_{kl}$ real and $\mu_{kl}=\mu_{lk}$. Then the relevant gate reads
\begin{equation}
J=\naw{V\otimes V}\widetilde{J}\naw{V^+\otimes V^+}.\label{a1}
\end{equation}
We see that $J$ depends on $N-1+\binom{N-1}{2}=\binom{N}{2}$ free parameters.

The above definition is quite general. In fact, it seems that this is the only freedom left if one assumes that all classical strategies are properly included is related to the choice of phases for the basic vectors (\ref{ab}).

\section{Maximally entangled games}
We call the ELW game maximally entangled if the initial state (\ref{a}) is maximally entangled. Let
\begin{equation}
\rho_i=\ket{\Psi_i}\bra{\Psi_i}\label{b}
\end{equation} 
be the density matrix corresponding to the initial state. The state described by $\rho_i$ is maximally entangled if the reduced density matrices are proportional to the unit matrix \cite{Nielsen}
\begin{equation}
\text{Tr}_A\rho_i=\frac{1}{N}I,\qquad \text{Tr}_B\rho_i=\frac{1}{N}I.\label{bb}
\end{equation} 
The maximally entangled game is distinguished by its properties. Consider first the $N=2$ case. The game can be described in terms of quaternion algebra \cite{Landsburg1} and real Hilbert space \cite{Bolonek}. What is more important, to any strategy of one player there exists an appropriate counterstrategy of the second player which leads to any outcome he/she desires \cite{BenjaminHay}, \cite{Landsburg1}. As a result no nontrivial pure Nash equilibrium exists while the form of the mixed one is strongly restricted \cite{Landsburg}.

The existence of counterstrategies in the case of maximally entangled game can be established for any $N$ \cite{Bolonek2}; it results in a simple way from the following property of such a game: the stability group of the initial state $\ket{\Psi_i}$ is (up to an automorphism) the diagonal subgroup of $SU(N)\times SU(N)$ \cite{Bolonek2}. Therefore, for any $N$ the maximally entangled games exhibit no nontrivial pure Nash equilibria. The structure of the mixed ones will be analyzed in a separate paper \cite{Bolonek4}. 

In the present section we derive the general conditions on the parameters $\lambda_k$ and $\mu_{kl}$ which yield the gate for maximally entangled game. To this end let us write out explicitly the initial density matrix $\rho_i$. With the help of equation (\ref{a1}) we find
\begin{equation}
\rho_i=\naw{V\otimes V}\widetilde{J}\naw{V^+\ket{1}\otimes V^+\ket{1}}\naw{\bra{1}V\otimes\bra{1}V}\widetilde{J}^+\naw{V^+\otimes V^+}.\label{b1}
\end{equation}
Let us note that $\widetilde{J}$ is diagonal and can be written as
\begin{equation}
\widetilde{J}_{\alpha\beta,\gamma\delta}=\widetilde{J}_{\alpha\beta}\delta_{\alpha\gamma}\delta_{\beta\delta}\label{b2}
\end{equation}
where $\widetilde{J}_{\alpha\beta}$ are the diagonal elements of $\widetilde{J}$. Using eqs. (\ref{bb}), (\ref{b1}) and (\ref{b2}) one easily finds that the condition $\text{Tr}_{B}\rho_i=\frac{1}{N}\text{I}$ can be written as
\begin{equation}
\frac{1}{N}\underset{\widetilde{}}{J}\underset{\widetilde{}}{J}^+=I\label{b3}
\end{equation}
where $\underset{\widetilde{}}{J}$ is the $N\times N$ matrix defined by
\begin{equation}
\underset{\widetilde{}}{J}{}_{\alpha\beta}=\widetilde{J}_{\alpha\beta}.
\end{equation}
The indices $\alpha$ and $\beta$ on the left hand side number the matrix elements of $\underset{\widetilde{}}{J}$ while on the right hand side - the diagonal elements of $\widetilde{J}$.\\
From eq. (\ref{b3}) we can derive a set of conditions on the parameters $\lambda_k$ and $\mu_{kl}$ defining the gate. To this end we select the basis in the Cartan subalgebra of $SU(N)$. A convenient choice is 
\begin{equation}
\naw{\Lambda_k}_{\alpha\beta}=\delta_{k\alpha}\delta_{\alpha\beta}-\delta_{k+1\alpha}\delta_{\alpha\beta}\label{db}
\end{equation}
$k=1,...,N-1$, $\alpha,\beta=1,...,N$. Inserting this expression into eq. (\ref{aa}) one finds the explicit form of the matrix  $\underset{\widetilde{}}{J}$ in terms of the parameters $\lambda_k$ and $\mu_{kl}$:
\begin{equation}
\begin{split}
& \underset{\widetilde{}}{J}{}_{\alpha\beta}=\exp\left( i\naw{\lambda_{\alpha}+\lambda_{\alpha-1}}\delta_{\alpha\beta}-i\lambda_\alpha\delta_{\alpha+1\beta}-i\lambda_{\alpha-1}\delta_{\alpha\beta+1}+\right.\\
&\qquad\quad\left. +i\naw{\mu_{\alpha\beta}+\mu_{\alpha-1\beta-1}-\mu_{\alpha-1\beta}-\mu_{\alpha\beta-1}}\right)
\end{split}\label{b4}
\end{equation}
where, by definition, $\lambda_0=\lambda_N=\mu_{0\alpha}=\mu_{\alpha 0}=\mu_{N\alpha}=\mu_{\alpha N}=0$.
Eqs. (\ref{b3}) and (\ref{b4}) provide the set of equations determining the parameters $\lambda_k$ and $\mu_{kl}$. In short, the necessary and sufficient condition for the gate operator $J$ to define a maximally entangled game is that the matrix $\frac{1}{\sqrt{N}}\underset{\widetilde{}}{J}$, with  $\underset{\widetilde{}}{J}$ defined by eq. (\ref{b4}), is unitary.

It is trivial to check with the help of eq. (\ref{b4}) the validity of diagonal part of eq. (\ref{b3}). So the only nontrivial conditions are obtained by demanding the vanishing of $\binom{N}{2}$ independent off-diagonal elements of $\underset{\widetilde{}}{J}\underset{\widetilde{}}{J}^+$. The relevant equations read
\begin{equation}
\sum_{\beta=1}^{N}e^{i\varphi_{\alpha\gamma;\beta}}=0,\qquad\alpha,\gamma=1,..,N,\qquad \alpha<\gamma\label{ab1}
\end{equation} 
with $\varphi_{\alpha\gamma;\beta}$ defined by
\begin{equation}
\begin{split}
& \varphi_{\alpha\gamma;\beta}=\naw{\lambda_\alpha+\lambda_{\alpha-1}}\delta_{\alpha\beta}-\naw{\lambda_\gamma+\lambda_{\gamma-1}}\delta_{\gamma\beta}+\\
& -\naw{\lambda_\alpha\delta_{\alpha\beta-1}-\lambda_\gamma\delta_{\gamma\beta-1}} -\naw{\lambda_{\alpha-1}\delta_{\alpha\beta+1}-\lambda_{\gamma-1}\delta_{\gamma\beta+1}}+\\
&+\naw{\mu_{\alpha\beta}-\mu_{\gamma\beta}+\mu_{\alpha-1\beta-1}-\mu_{\gamma-1\beta-1}-\mu_{\alpha-1\beta}+\mu_{\gamma-1\beta}-\mu_{\alpha\beta-1}+\mu_{\gamma\beta-1}}(\text{mod}2\pi).
\end{split}\label{ab2}
\end{equation}
Let us note that
\begin{equation}
\sum_{\beta=1}^{N}\varphi_{\alpha\gamma;\beta}=0 \quad(\text{mod}2\pi).\label{ab3}
\end{equation}
Eqs. (\ref{ab1}) and (\ref{ab2}) provide the general solution to our problem. The convenient strategy to solve them is to find first the solutions to the eqs. (\ref{ab1}) and (\ref{ab3}) which yield the values of $\varphi_{\alpha\gamma;\beta}$ and then to solve eqs. (\ref{ab2}) in terms of $\lambda_\alpha$ and $\mu_{\alpha\beta}$. Some examples of the solutions are provided in the next section.

\section{Some examples}
The case $N=2$ corresponds to $SU(2)$ group. The Cartan subalgebra is onedimensional and is spanned by
\begin{equation}
\Lambda=\sigma_3.
\end{equation}
There is one parameter $\lambda$ and eq. (\ref{b3}) yields
\begin{equation}
\underset{\widetilde{}}{J}=\left(\begin{array}{cc}
e^{i\lambda} & e^{-i\lambda}\\
e^{-i\lambda} & e^{i\lambda} \end{array}\right).
\end{equation}
Eq. (\ref{b3}) leads to the following condition
\begin{equation}
e^{4i\lambda}+1=0
\end{equation}
i.e. $\lambda=\frac{\pi}{4}$.

Consider now the case $N=3$. There are two basic elements of Cartan subalgebra of $SU(3)$:
\begin{equation}
\Lambda_1=\left(\begin{array}{ccc}
1 & 0 & 0\\
0 & -1 & 0\\
0 & 0 & 0 \end{array}\right),\qquad \Lambda_2=\left(\begin{array}{ccc}
0 & 0 & 0\\
0 & 1 & 0\\
0 & 0 & -1 \end{array}\right)
\end{equation}
and three parameters $\lambda_1$, $\lambda_2$ and $\mu_{12}=\mu_{21}$. Eqs. (\ref{b3}) and (\ref{b4}) yield a set of conditions which are equivalent to those derived in Ref. \cite{Bolonek1} (one has only to take into account the different choice of the Cartan subalgebra basis in \cite{Bolonek1}). It is shown in \cite{Bolonek1} that the resulting equations admit a discrete set of solutions. This can be explained as follows. Eq. (\ref{ab1}) takes the form
\begin{equation}
e^{i\varphi_1}+e^{i\varphi_2}+e^{i\varphi_3}=0
\end{equation}
By multiplying both sides by $\exp\naw{-i\varphi_1}$ we conclude that, up to renumbering, $\varphi_2-\varphi_1=\frac{2\pi}{3}$, $\varphi_3-\varphi_1=\frac{4\pi}{3}$. Therefore, we find
\begin{equation}
\begin{split}
& \varphi_1=\varphi\\
& \varphi_2=\varphi+\frac{2\pi}{3}\\
& \varphi_3=\varphi+\frac{4\pi}{3}.
\end{split}
\end{equation}
Now, eq. (\ref{ab3}) impies that $\varphi_1$, $\varphi_2$ and $\varphi_3$, modulo $2\pi$ and up to a renumbering, equal $0$, $\frac{2\pi}{3}$ and $\frac{4\pi}{3}$, respectively. This yields, via eqs. (\ref{ab2}), the discrete set of solutions. We omit the details here.

Let us pass to the case $N=4$. The gate operator is now parametrized by six quantities $\lambda_1$, $\lambda_2$, $\lambda_3$, $\mu_{12}=\mu_{21}$, $\mu_{13}=\mu_{31}$ and $\mu_{23}=\mu_{32}$. According to the eqs. (\ref{b3}) and (\ref{b4}) demanding the maximal entanglement is equivalent to the unitarity of the matrix:
\begin{equation}
\frac{1}{\sqrt{N}}\underset{\widetilde{}}{J}=\frac{1}{2}\left(\begin{array}{cccc}
e^{i\lambda_1} & e^{i\naw{-\lambda_1+\mu_{12}}} & e^{i\naw{\mu_{13}-\mu_{12}}} & e^{-i\mu_{13}}\\
e^{i\naw{-\lambda_1+\mu_{12}}} & e^{i\naw{\lambda_1+\lambda_2-2\mu_{12}}} & e^{i\naw{-\lambda_2+\mu_{23}+\mu_{12}-\mu_{13}}} & e^{i\naw{\mu_{13}-\mu_{23}}}\\
e^{i\naw{\mu_{13}-\mu_{12}}} & e^{i\naw{-\lambda_2+\mu_{12}+\mu_{23}-\mu_{13}}} & e^{i\naw{\lambda_2+\lambda_3-2\mu_{23}}} & e^{i\naw{-\lambda_3+\mu_{23}}}\\
e^{-i\mu_{13}} & e^{i\naw{\mu_{13}-\mu_{23}}} & e^{i\naw{-\lambda_3+\mu_{23}}} & e^{i\lambda_3} \end{array}\right).
\end{equation}
Eqs. (\ref{ab1}) and (\ref{ab2}) take now the following form
\begin{equation}
e^{i\naw{2\lambda_1-\mu_{12}}}+e^{i\naw{-2\lambda_1-\lambda_2+3\mu_{12}}}+e^{i\naw{\lambda_2+2\mu_{13}-2\mu_{12}-\mu_{23}}}+e^{i\naw{-2\mu_{13}+\mu_{23}}}=0\label{c1}
\end{equation}
\begin{equation}
e^{i\naw{\lambda_1-\mu_{13}+\mu_{12}}}+e^{i\naw{-\lambda_1+\lambda_2+\mu_{13}-\mu_{23}}}+e^{i\naw{-\lambda_2-\lambda_3+2\mu_{23}-\mu_{12}+\mu_{13}}}+e^{i\naw{\lambda_3-\mu_{23}-\mu_{13}}}=0
\end{equation}
\begin{equation}
e^{i\naw{\lambda_1+\mu_{13}}}+e^{i\naw{-\lambda_1+\mu_{12}-\mu_{13}+\mu_{23}}}+e^{i\naw{\lambda_3-\mu_{23}+\mu_{13}-\mu_{12}}}+e^{i\naw{-\mu_{13}-\lambda_3}}=0
\end{equation}
\begin{equation}
e^{i\naw{-\lambda_1+2\mu_{12}-\mu_{13}}}+e^{i\naw{\lambda_1+2\lambda_2-3\mu_{12}-\mu_{23}+\mu_{13}}}+e^{i\naw{-2\lambda_2-\lambda_3+3\mu_{23}+\mu_{12}-\mu_{13}}}+e^{i\naw{\lambda_3-2\mu_{23}+\mu_{13}}}=0
\end{equation}
\begin{equation}
e^{i\naw{-\lambda_1+\mu_{12}+\mu_{13}}}+e^{i\naw{\lambda_1+\lambda_2-2\mu_{12}-\mu_{13}+\mu_{23}}}+e^{i\naw{-\lambda_2+\lambda_3+\mu_{12}-\mu_{13}}}+e^{i\naw{\mu_{13}-\mu_{23}-\lambda_3}}=0
\end{equation}
\begin{equation}
e^{i\naw{2\mu_{13}-\mu_{12}}}+e^{i\naw{-\lambda_2+2\mu_{23}+\mu_{12}-2\mu_{13}}}+e^{i\naw{\lambda_2+2\lambda_3-3\mu_{23}}}+e^{i\naw{-2\lambda_3+\mu_{23}}}=0\label{c2}
\end{equation}
Generically, we have:
\begin{equation}
e^{i\varphi_1}+e^{i\varphi_2}+e^{i\varphi_3}+e^{i\varphi_4}=0\label{c3}
\end{equation}
\begin{equation}
\varphi_1+\varphi_2+\varphi_3+\varphi_4=0.\label{c4}
\end{equation}
By multiplying both sides of eq. (\ref{c3}) by $\exp\naw{-\frac{i}{2}\naw{\varphi_1+\varphi_2}}$ we get
\begin{equation}
e^{-\frac{i}{2}\naw{\varphi_2-\varphi_1}}+e^{\frac{i}{2}\naw{\varphi_2-\varphi_1}}+e^{-i\naw{\varphi_3-\frac{\varphi_1+\varphi_2}{2}}}+e^{-i\naw{\varphi_4-\frac{\varphi_1+\varphi_2}{2}}}=0.
\end{equation}
Thus, up to a renumbering
\begin{equation}
\begin{split}
& \varphi_3=\varphi_1+\pi(\text{mod}2\pi)\\
& \varphi_4=\varphi_2+\pi(\text{mod}2\pi).
\end{split}
\end{equation}
Eq. (\ref{c4}) yields
\begin{equation}
\varphi_1+\varphi_2=0(\text{mod}\pi)
\end{equation}
and we arrive at the one-parameter family of solutions which can be conveniently parametrized as
\begin{equation}
\varphi_2=n\pi-\varphi_1\label{c5}
\end{equation}
\begin{equation}
\varphi_3=\varphi_1+\naw{2m+1}\pi
\end{equation}
\begin{equation}
\varphi_4=-\varphi_1+(2m+n+1)\pi\label{c7}
\end{equation}
 where $m$ and $n$ are arbitrary integers. Let us note that we obtain, for fixed $m$ and $n$, the one parameter family of solutions.
 
 For any of eqs. (\ref{c1})$\div$(\ref{c2}) the variables $\varphi_i$ are different functions of the initial parameters $\lambda_k$ and $\mu_{kl}$. By combining different solutions (\ref{c5})$\div$(\ref{c7}) one obtains numerous solutions to the original equations (\ref{c1})$\div$(\ref{c2}). We shall quote below few of them. An important point is that, contrary to the $N=2$ and $N=3$ cases, we are dealing here with the one parameter families of solutions. It is clear from the derivation given above that one can expect the existence of such solutions for all $N\geq 4$; moreover, the number of free parameters will grow.
 
  Some solutions of equations (\ref{c1})$\div$(\ref{c2}) are listed below:  
 
\begin{equation}\begin{split}
& \left\{\begin{array}{c}
\lambda_1=-\beta-\pi+\pi p\\
\lambda_2=2\mu_{23}-\pi\\
\lambda_3=-\beta+\pi p\\
\mu_{12}=\mu_{23}-\pi\\
\mu_{13}=\beta+\mu_{23}\\
\mu_{23}=\frac{\pi}{2},\frac{3\pi}{2},\frac{\pi}{2}-2\beta,\frac{3\pi}{2}-2\beta \end{array}\right.
\qquad
\left\{\begin{array}{c}
\lambda_1=-\beta+\frac{\pi}{4}+\pi p\\
\lambda_2=2\mu_{23}-\pi\\
\lambda_3=-\beta+\frac{\pi}{4}+\pi p\\
\mu_{12}=\mu_{23}-\frac{\pi}{2}\\
\mu_{13}=\beta+\mu_{23}\\
\mu_{23}=\frac{3\pi}{4},\frac{7\pi}{4},-2\beta,\pi-2\beta \end{array}\right.\\
& \left\{\begin{array}{c}
\lambda_1=-\beta-\frac{\pi}{2}+\pi p\\
\lambda_2=2\mu_{23}-\pi\\
\lambda_3=-\beta+\frac{\pi}{2}+\pi p\\
\mu_{12}=\mu_{23}\\
\mu_{13}=\beta+\mu_{23}\\
\mu_{23}=0,\pi,\frac{\pi}{2}-2\beta,\frac{3\pi}{2}-2\beta \end{array}\right.
\qquad
\left\{\begin{array}{c}
\lambda_1=-\beta+\frac{3\pi}{4}+\pi p\\
\lambda_2=2\mu_{23}-\pi\\
\lambda_3=-\beta+\frac{3\pi}{4}+\pi p\\
\mu_{12}=\mu_{23}+\frac{\pi}{2}\\
\mu_{13}=\beta+\mu_{23}\\
\mu_{23}=\frac{\pi}{4},\frac{5\pi}{4},-2\beta,\pi-2\beta \end{array}\right.\\
& \left\{\begin{array}{c}
\lambda_1=\beta+\pi p+\pi q\\
\lambda_2=\frac{\pi}{2}q\\
\lambda_3=\beta+\pi p\\
\mu_{12}=\frac{3\pi}{4}q+\pi\\
\mu_{13}=\beta+\frac{\pi}{4}q\\
\mu_{23}=\frac{\pi}{4}q \end{array}\right.
\qquad
\left\{\begin{array}{c}
\lambda_1=-\beta+\frac{5\pi}{4} p-\frac{\pi}{2} q\\
\lambda_2=-4\beta+\pi p\\
\lambda_3=-\beta+\frac{\pi}{4} p-\frac{\pi}{2}q\\
\mu_{12}=-2\beta+\pi+\pi p+\pi q\\
\mu_{13}=-\beta+\frac{\pi}{2}p\\
\mu_{23}=-2\beta+\frac{\pi}{2}p \end{array}\right.
\end{split}
\end{equation}
where $\beta$ is a real parameter; $p$ and $q$ are integers.

\section{Solution for general N}

It is quite easy to find the general form of the gate  leading to maximally entangled game without solving eqs. (\ref{ab1}) and (\ref{ab2}). The price we have to pay is that some games are then equivalent in the sense described below. \\
According to the eqs. (\ref{b2}) and (\ref{b3}) the general form of the gate  under consideration reads
\begin{equation}
\widetilde{J}_{\alpha\beta,\gamma\delta}=\sqrt{N}W_{\alpha\beta}\delta_{\alpha\gamma}\delta_{\beta\delta}\label{ca}
\end{equation}
where $W$ is some unitary matrix, $WW^+=I$. However, $W$ cannot be completely arbitrary. In order to preserve the symmetry of the game one must impose
\begin{equation}
W_{\alpha\beta}=W_{\beta\alpha}\label{ca1}
\end{equation}
so $W$ may be arbitrary unitary symmetric matrix. Deleting the trivial overall factor we assume that $W$ is a symmetric element of $SU(N)$. Such elements are generated by $N-1$ traceless diagonal real matrices and $\binom{N}{2}$ off-diagonal real symmetric ones. On the level of the gate matrix $\widetilde{J}$ this implies that we should supply $\binom{N}{2}$ generators entering the exponent on the right hand side of eq. (\ref{aa}) by $N-1$ generators of the form $I\otimes \Lambda_k+\Lambda_k\otimes I$, $k=1,...,N-1$. However, denoting by $\widetilde{J}\,^{'}$ a new gate  one easily finds with the help of BCH formula
\begin{equation}
\widetilde{J}\,^{'}=\naw{\widetilde{J}_A\otimes\widetilde{J}_B}\widetilde{J}.
\end{equation}
Therefore, by relabelling the strategies,
\begin{equation}
U_A\rightarrow \widetilde{J}_A  U_A\widetilde{J}^+_A,\qquad U_B\rightarrow \widetilde{J}_B U_B\widetilde{J}^+_B
\end{equation}
we arrive at the game differing only by classification of strategies. 

Neglecting the above subtlety we conclude that the general gate operator leading to maximally entangled game is given by eqs. (\ref{a1}), (\ref{ca}) and (\ref{ca1}).

Mathematically, the games differing only by relabelling of strategies equivalent in the sense that any question concerning one game (for example, the localization and classification of Nash equilibria, the existence of Pareto optimally strategies etc.) can be easily translated into equivalent question concerning the second game. However, when one considers the physical realization of the game (like, for example in Ref. \cite{DuLi1} where the game is implemented using two qubit nuclear magnetic resonance quantum computer) the games are no longer equivalent in the sense that the gate takes a particular form determined by the underlying physical mechanism (the same concerns the implementation of strategies). Referring to the above example described in \cite{DuLi1} the single parameter characterizing the gate is determined, among others, by the spin-spin couplings between the nuclei.

\section{Concluding remarks}

We have considered the maximally entangled 2-players N-strategies games. The necessary and sufficient conditions for the game defined by the gate (\ref{a1}) to be maximally entangled are given by eqs. (\ref{b3}) and (\ref{b4}).

In order to find their explicit form we have to solve equations (\ref{ab1}) and (\ref{ab2}). The most convenient choice is to find first the general solution to the equations (\ref{ab1}) and (\ref{ab3}) and then to solve eqs. (\ref{ab2}) in terms of $\lambda$'s and $\mu$'s. In this way we obtain a rich variety of solutions. The important point is that only for N=2 and 3 these solutions form the discrete sets. Starting from N=4 we obtain the continuous families of solutions with growing number of free parameters.

The admissible gate  $\widetilde{J}$ may be multiplied by the product of arbitrary unitary matrices $\widetilde{J}_{A,B}$ leading to new gate operator
\begin{equation}
\widetilde{J}\,^{'}=\naw{\widetilde{J}_A\otimes\widetilde{J}_B}\widetilde{J}.
\end{equation}
This amounts only to relabelling of Alice and Bob strategies $U_A\rightarrow \widetilde{J}_AU_A\widetilde{J}^+_A$,  $U_B\rightarrow \widetilde{J}_B U_B\widetilde{J}^+_B$. If we, however, does distinguish between such games the construction of relevant gate  becomes particularly simple (cf. eqs. (\ref{a1}), (\ref{ca}) and (\ref{ca1})).

All games considered above are symmetric with respect to the exchange of players. This is a natural situation from the point of view of game theory. However, we can admit a more general situation that, on the classical level, the number of strategies at Alice disposal, $M$, differs from the number of Bob strategies, $N$; without loosing generality one can assume $M\geq N$. The classical payoffs are now defined by $M\times N$ matrices. At the quantum level we ascribe to Alice (Bob) the $M$-dimensional ($N$-dimensional) complex Hilbert space $\mathcal{H}^{(A)}$ ($\mathcal{H}^{(B)}$), respectively; the total Hilbert space of the game is now $\mathcal{H}^{(A)}\otimes \mathcal{H}^{(B)}$. The starting point is again the vector $\ket{1}_A\otimes \ket{1}_B$ (in, slightly generalized, notation introduced in eq. (\ref{ab})). There is now only one condition on the gate $J$ (the symmetry condition must be abandoned): all classical pure strategies are contained in the set of pure quantum ones. One can then construct $J$ along the similar lines as in the symmetric case. The result is that $J$ is locally unitarily equivalent to the unitary matrix constructed in terms of tensor products of elements of Cartan subalgebras of $SU(M)$ and $SU(N)$ (cf. eqs. (\ref{aa}) and (\ref{a1})). The condition for maximal entanglement reads now \cite{Nielsen}
\eq{\text{Tr}_A\rho_i=\frac{1}{N}\cdot I}
with
\eq{\rho_i=\ket{\Psi_i}\bra{\Psi_i},\qquad \ket{\Psi_i}=J\naw{\ket{1}_A\otimes\ket{1}_B}.}
One can now proceed as in the symmetric case arriving at the equations generalizing eqs.  (\ref{ab1})$\div$(\ref{ab3}).

As we have pointed out above the symmetric case seems to more natural from game-theoretic point of view. However, the asymmetric case is still very interesting. Using the method presented in \cite{Bolonek5} and applying Schur's lemma one can find the stability subgroup of the initial state corresponding to the maximal entanglement. It has quite nontrivial (although simple) structure which (as we discuss briefly in Sec. II) strongly influences the properties of the game. Therefore, the asymmetric case is worth of being studied in more detail.

\subsection*{Acknowledgment}
Research of Katarzyna Bolonek-Laso\'n was supported by NCN Grant\\ no. DEC-2012/05/D/ST2/00754. Piotr Kosi\'nski gratefully acknowledges the fruitful discussion with P. Ma\'slanka, K. Andrzejewski and J. and C. Gonera. We thank anonymous referee for interesting and useful remarks.

\begin{appendices}

\section{Cartan subalgebras \cite{Hump}}
We do not need the general definition of Cartan subalgebra for arbitrary Lie algebra. In the case of semisimple Lie algebras over the complex numbers the Cartan subalgebras are uniquely defined by the following properties:
\begin{itemize}
\item[(i)] they are maximal abelian subalgebras 
\item[(ii)] if an element $X$ belongs to the Cartan subalgebra then ad$X$ is diagonalisable.
\end{itemize}
All Cartan subalgebras are related by inner automorphisms.
The simultaneous diagonalization of all endomorphisms ad$X$ with $X$ running over the basis of some Cartan subalgebra allows to give the full characterization of the structure of Lie algebra under consideration. In particular, the eigenspace corresponding to zero eigenvalue coincides with the Cartan subalgebra itself while the remaining eigenspaces are onedimensional.\\
The elements of Lie algebra of $SU(N)$ group are the traceless hermitean $N\times N$ matrices. By virtue of (i) they commute so they can be diagonalised simultaneously. Therefore, the maximal set of hermitean traceless diagonal matrices forms Cartan subalgebra of $SU(N)$ algebra. The basis of this subalgebra can be chosen as in eq. (\ref{db}).

\end{appendices}

\end{document}